%% file: flaglets_sampta_2013.tex
\documentclass[a4paper,conference]{IEEEtran}
\usepackage{amsmath,epsfig}
\usepackage{amsfonts}
\usepackage{amssymb}
\usepackage{graphicx} 
\usepackage{color} 
\usepackage{subfigure} 
\usepackage{ifpdf}
\ifpdf
  \usepackage[pdftex]{hyperref}
\else
  \usepackage[ps2pdf,colorlinks=true]{hyperref}
\fi

\input{macros}

\renewcommand{\eqn}[1]{Eqn.~(#1)}

\renewcommand{\exp}[1]{\ensuremath{{{\rm exp}({#1})}}}

\makeatletter
\newcommand\footnoteref[1]{\protected@xdef\@thefnmark{\ref{#1}}\@footnotemark}
\makeatother

\begin{document}\sloppy

\title{Fourier-Laguerre transform, Convolution and Wavelets on the Ball}

\author{%
\IEEEauthorblockN{
Jason D. McEwen\IEEEauthorrefmark{1} and
Boris Leistedt\IEEEauthorrefmark{1}} 
\IEEEauthorblockA{\IEEEauthorrefmark{1} 
Department of Physics and Astronomy, University College London, London WC1E 6BT, U.K.\\
\tt\footnotesize \{jason.mcewen, boris.leistedt.11\}@ucl.ac.uk}
}

\maketitle

\begin{abstract}
  We review the Fourier-Laguerre transform, an alternative harmonic
  analysis on the three-dimensional ball to the usual Fourier-Bessel
  transform.  The Fourier-Laguerre transform exhibits an exact
  quadrature rule and thus leads to a sampling theorem on the ball.
  We study the definition of convolution on the ball in this context,
  showing explicitly how translation on the radial line may be viewed
  as convolution with a shifted Dirac delta function.  We review the
  exact Fourier-Laguerre wavelet transform on the ball, coined
  \emph{flaglets}, and show that flaglets constitute a tight frame.
\end{abstract}


% -------------------------------------------------------------------------
\section{Introduction}
\label{sec:intro}

Data often live naturally on the three-dimensional ball.  For example,
in cosmology the distribution of galaxies that traces the large-scale
structure of the Universe is observed on the celestial sphere (\eg\
\cite{ahn:2012}), augmented with depth information given by
redshift. A spherical shell at a given redshift represents a given
epoch in the history of our Universe; thus, such data live naturally
on the three-dimensional ball (hereafter referred to as simply the
\emph{ball}).

One would like to analyse such data-sets on the ball to study the
physics responsible for them.  Since many physical processes are
manifest on different physical scales, while also spatially localised,
wavelet analysis is a power method for this purpose.  Recently, two
wavelet transforms have been derived on the ball
\cite{lanusse:2012,leistedt:flaglets}.  The former \cite{lanusse:2012}
is based on an undecimated wavelet construction, built on the
Fourier-Bessel transform.  The latter \cite{leistedt:flaglets} is
based on a tiling of harmonic space, built on a Fourier-Laguerre
transform, and developed by the authors of the current article.  Our
approach \cite{leistedt:flaglets}: (i) yields wavelets that are not
isotropic but rather exhibit an angular opening that is invariant
under radial translation; (ii) is theoretically exact; and (iii) leads
to a fast multiresolution algorithm.

In this article we review our recent work \cite{leistedt:flaglets}
where we consider the Fourier-Laguerre transform and construct
wavelets (which we coin \emph{flaglets}) on the ball.  Furthermore, we
illuminate the translation operator on the radial line, showing
how this may be viewed as convolution with a shifted Dirac delta
function.  We also show that flaglets constitute a tight frame.

% -------------------------------------------------------------------------
\section{Fourier-Laguerre transform}
\label{sec:flag}

The canonical harmonic transform on the ball is the Fourier-Bessel
transform, where the basis functions are the eigenfunctions of the
Laplacian on the ball.  The Fourier-Bessel basis functions separate
into the usual spherical harmonic functions on the sphere and the
spherical Bessel functions on the radial line.  However, the
Fourier-Bessel transform suffers from a serious shortcoming.  To
the best of our knowledge there does not exist a sampling theorem for
the Fourier-Bessel transform, since there does not exist an exact
quadrature rule for the evaluation of the spherical Bessel transform
(the radial part of the Fourier-Bessel transform).

To overcome this limitation we consider the Fourier-Laguerre
transform, for which we developed a sampling theorem
\cite{leistedt:flaglets}.  The Fourier-Laguerre transform follows by
adopting the Laguerre polynomials (the standard orthogonal polynomials
on $\reals^{+}$) as the radial basis functions, while keeping the
spherical harmonics as the spherical basis functions. We define the
Fourier-Laguerre basis functions on the ball \mbox{$\ball=
  {\mathbb{R}^+} \times \sphere$} by
\begin{equation}
  Z_{\ell \m \ip}(\vect{r}) = K_{\ip}(\rad)
  \shfarg{\ell}{\m}{\sas} ,
\end{equation}
with spherical coordinates \mbox{$\vect{r} = (r, \sas) \in \ball$}, where
$\rad\in\reals^{+}$ denotes radius, $\saa\in[0,\pi]$ colatitude and
$\sab\in[0,2\pi)$ longitude, and where $\el,\ip\in\naturals_0$ and
$\m\in\integers$ such that $|m|\leq\el$. The standard spherical
harmonics are denoted by $\shf{\ell}{\m}$ and the normalised spherical
Laguerre basis functions are defined on the radial line by
\begin{equation}
  K_\ip(r) \equiv \sqrt{ \frac{\ip!}{(\ip+2)!} }  
    \frac{ e^{-{r}/{2\tau}} }{ \sqrt{\tau^3}} L^{(2)}_\ip\left({r /
        \tau}\right),
\end{equation}
where $L^{(2)}_\ip$ is the $\ip$-th generalised Laguerre polynomial of
order two and $\tau \in \mathbb{R}^+$ is a radial scale factor. 

A square-integrable signal $f \in \ltwo(\ball)$ can then be decomposed as
\begin{equation}
\label{eqn:flag_inverse}
  f(\vect{\rad}) = 
  \sum_{\ip = 0}^{\infty}\sum_{\el = 0}^{\infty}\sum_{\m = -\el}^{\el} 
  f_{\el \m \ip} Z_{\el \m \ip} (\vect{\rad}),
\end{equation}
where the harmonic coefficients are given by the usual projection
\begin{equation}
\label{eqn:flag_forward}
  {f}_{\el \m \ip} = \langle f | Z_{\el \m \ip} \rangle_\ipball
  =  \int_{\ball} \dx^3\vect{\rad}  f(\vect{\rad}) Z^{*}_{\el \m \ip}(\vect{\rad}),
\end{equation}
where $\dx^3\vect{r} = \rad^2 \sin\saa \dx r \dx \saa \dx \sab$ is the
usual rotation invariant measure in spherical
coordinates.\footnote{This measure is a natural choice since it allows
  the Fourier-Laguerre transform to be related directly to the
  Fourier-Bessel transform, such that the Fourier-Bessel coefficients
  can be computed exactly from Fourier-Laguerre coefficients
  (see \cite{leistedt:flaglets} for further details).}
%%%
% Following footnote remove due to space constraints.
% \footnote{Note that the radial component of the
%   Fourier-Laguerre transform, namely the spherical Laguerre transform,
%   is a real transform, which can be extended to complex signals by
%   considering their real and imaginary parts separately.} 
%%%
We consider band-limited signals, with angular and radial band-limits
$L$ and $P$, respectively, \ie\ signals $f$ such that \mbox{$f_{\el \m
    \ip} = 0$}, $\forall \el \geq \elmax$, $\forall \ip \geq
\ipmax$. In this case the summations in \eqn{\ref{eqn:flag_inverse}}
over \el\ and \ip\ may be truncated to $\elmax-1$ and $\ipmax-1$
respectively.

In practice, computing the Fourier-Laguerre transform involves the
evaluation of the integral of \eqn{\ref{eqn:flag_forward}}. An exact
quadrature rule for the evaluation of this integral for a band-limited
function $f$ naturally gives rise to a sampling theorem.  Since the
Fourier-Laguerre transform is separable in angular and radial
coordinates, we may appeal to separate sampling theorems on the sphere
and radial line.  For the angular part, we adopt the equiangular
sampling theorem on the sphere developed recently by one of the
authors \cite{mcewen:fssht}.  Other sampling theorems on the sphere
could alternatively be adopted (\eg\ \cite{driscoll:1994}), however we
select the sampling theorem developed by \cite{mcewen:fssht} since it
leads to the most efficient sampling of the sphere (\ie\ the fewest
number of samples to represent a band-limited signal exactly).  For
the radial part, we appeal to Gaussian quadrature to develop an exact
quadrature rule and, consequently, a sampling theorem
\cite{leistedt:flaglets}.  Combining these results we recover a
sampling theorem and, equivalently, an exact Fourier-Laguerre
transform on $\ball$.  For a band-limited signal all of the
information content of the signal is captured in $N =
\ipmax[(2\elmax-1)(\elmax-1)+1] \sim 2 \ipmax \elmax^2$ samples on the
ball \cite{leistedt:flaglets}.

We have developed the public {\tt
  FLAG}\footnote{\url{http://www.flaglets.org/}} code
\cite{leistedt:flaglets} to compute the Fourier-Laguerre transform.
The {\tt FLAG} code computes exact forward and inverse
Fourier-Laguerre transforms at machine precision and is stable to
extremely large band-limits, relying on the public {\tt
  SSHT}\footnote{\url{http://www.jasonmcewen.org/}} code
\cite{mcewen:fssht} developed by one of the authors for the angular
part, which in turn relies on {\tt
  FFTW}\footnote{\url{http://www.fftw.org/}}.  {\tt FLAG} supports both
the C and Matlab programming languages.

% -------------------------------------------------------------------------
\section{Convolution on the ball}
\label{sec:convolution}

We review the definition of convolution on the ball
\cite{leistedt:flaglets}, highlighting how the translation operator
defined on the radial line may be viewed as convolution with a
Dirac delta function.
By the angular and radial separability of the Fourier-Laguerre
transform, we construct a convolution operator on the ball from
convolution operators on the sphere and radial line (\eg\
\cite{gorlich:1982}).

On the sphere, we adopt the usual convolution of $f \in
\ltwo(\sphere)$ with an axisymmetric kernel $h \in \ltwo(\sphere)$
given by the inner product (\eg\ \cite{wiaux:2007:sdw})
%\begin{equation}
%\begin{split}
\begin{align}
\label{eqn:spherical_convolution_real}
  (f \star h)(\sas) 
  &\equiv \langle f| \rot_{(\sas)}h\rangle_\ipsphere \\
  &= \int_{\sphere}
  \dmu{\saa\p,\sab\p} f(\saa\p,\sab\p) \left( \rot_{(\saa,\sab)} h
  \right)^\cconj(\saa\p,\sab\p), \nonumber
\end{align}
%\end{split}
%\end{equation}
where $\dmu{\sas} = \sin\saa \dx \saa \dx \sab$ is the usual rotation
invariant measure on the sphere. The translation operator on the
sphere is given by the standard three-dimensional rotation:
$(\rot_{(\euls)} h)(\sas) = h(\rot_{(\euls)}^{-1}(\sas))$, with
$(\euls) \in \sothree$, where $\eula\in[0,2\pi)$, $\eulb\in[0,\pi]$
and $\eulc\in[0,2\pi)$.  We make the association $\saa=\eulb$ and
$\sab=\eula$, \ie\ $\rot_{(\saa,\sab)} \equiv
\rot_{(\eula,\eulb,0)}$, and restrict our attention to convolution
with axisymmetric functions that are invariant under azimuthal rotation,
\ie\ $\rot_{(0,0,\eulc)} h = h$, so that we recover a convolved
function $f \star h$ defined on the sphere. In harmonic space,
axisymmetric convolution may be written 
\begin{equation}
  \shc{(f \star h)}{\el}{\m} = 
  \langle f \star h | \shf{\el}{\m} \rangle_\ipsphere 
  = \sqrt{ \frac{4\pi}{2\el+1}} \shc{f}{\el}{\m} \shcc{h}{\el}{0},
\end{equation}
with $\shc{f}{\el}{\m} = \langle f| \shf{\el}{\m} \rangle_\ipsphere$ and
$\shc{h}{\el}{0}\kron{\m}{0} = \langle h|\shf{\el}{\m} \rangle_\ipsphere$. The
generalisation to directional convolution on the sphere is
straightforward (see \eg\ \cite{wiaux:2007:sdw}), however we do not present it
here since we consider axisymmetric wavelets subsequently.

On the radial line, we consider a convolution operator appropriate for the
spherical Laguerre basis.  We adopt a convolution similar
to that considered by \cite{gorlich:1982} and others (see additional
references contained in \cite{gorlich:1982}), although we recover this
operator in an alternative manner.  Firstly, we define a translation
operator $\transl$ on the radial line, which is constructed by analogy
with the case for the infinite line, for which the standard orthogonal
basis is given by the complex exponentials $\phi_\omega(x) = \exp{{\rm
    i} \omega x}$, with $x,\omega \in \reals$.  Translation of the
basis functions on the infinite line is simply defined by the shift of
coordinates:
%\begin{equation}
$
  (\transl_{u}^\reals \phi_\omega)(x) 
  \equiv \phi_\omega(x-u) 
  = \phi_\omega^\cconj(u) \phi_\omega(x),
$ %
%\end{equation}
with $u\in\reals$ and where the final equality follows by the
standard rules for exponents.  We define translation of the spherical
Laguerre basis functions on the radial line by analogy:
\begin{equation}
  (\transl_{s} K_\ip)(\rad) \equiv K_\ip(s) K_\ip(\rad) ,
\end{equation}
where $s \in \reals^{+}$ (since $K_\ip$ is real we drop the
complex conjugation).  This leads to a natural harmonic expression for the
translation of a radial function $f\in\ltwo(\reals^+)$:
\begin{equation}
  \label{eqn:translation_radial_harmonic_expansion}
  (\transl_{s}f)(\rad) = 
  \sum_{\ip = 0}^{\infty}
  f_{\ip}  K_\ip(s) K_{\ip} (\rad) ,
\end{equation}
implying 
\begin{equation}
  \label{eqn:translation_radial_harmonic}
  {(\transl_{s}f)}_\ip = K_\ip(s) f_{\ip} ,
\end{equation}
where  ${f}_\ip = \langle f| K_\ip \rangle_\iphalfline$. 

With a translation operator to hand, we may define convolution on the
radial line of $f, h\in \ltwo(\reals^+)$ by the inner product
\begin{equation}
  (f \star h)(\rad) 
  \equiv \langle f | \transl_\rad h \rangle_\iphalfline
  =\int_{\reals^+}
  \dx s s^2 f(s) \left( \transl_{\rad} h \right)(s),
\end{equation}
from which it follows that radial convolution in harmonic space is
given by the product
\begin{equation}
  {(f \star h)}_{\ip} =
  \langle f \star h | K_\ip \rangle_\iphalfline 
  = f_\ip h_\ip,
\end{equation}
where  ${h}_\ip = \langle h| K_\ip \rangle_\iphalfline$. 

Although the definition of the convolution operator on the radial line
is complete, we would like to gain further intuition.  The action of
the translation operator is described in harmonic space through
\eqn{\ref{eqn:translation_radial_harmonic}}, which remains somewhat
opaque.  We would also like to view the translation operator that we
have constructed on the radial line in real space.

\begin{figure}
\centering
\setlength{\unitlength}{.5in}
\begin{picture}(9.5,2.1)(0,0)
\put(0.05,0.){\includegraphics[trim= 2.5cm 0cm 0.5cm 0cm, clip, width=9.4cm]{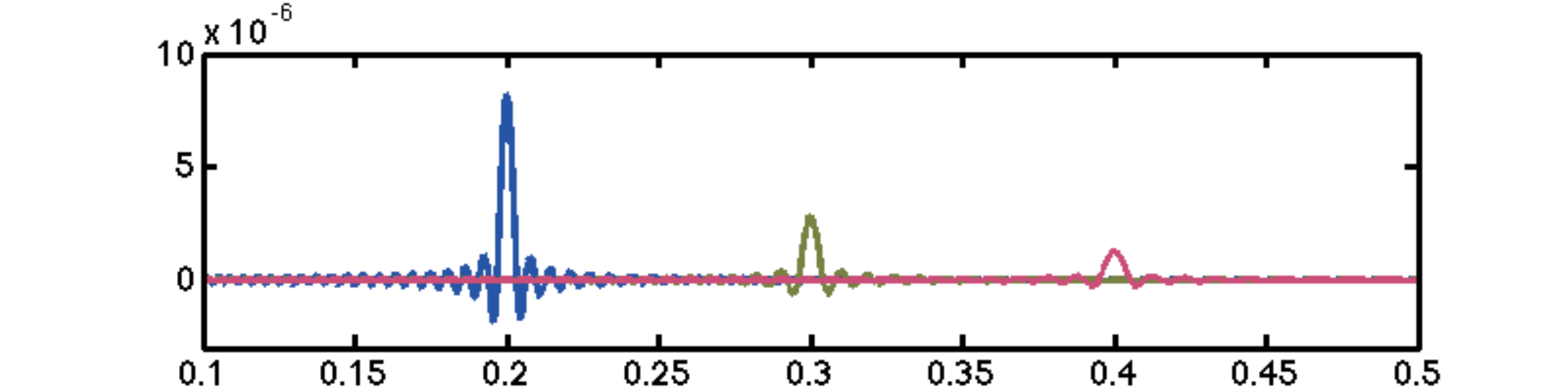}}
\put(3.6,-0.2){\scriptsize$r$}
\put(0.0,0.6){\rotatebox{90}{\scriptsize Amplitude}}
\end{picture}
\caption{Band-limited Dirac delta functions plotted on the radial line
  at positions $s=\{0.2, 0.3, 0.4\}$ (plotted in blue, green and red,
  respectively). Oscillations are caused by the finite band-limit
  (here $P=256$); as $P\rightarrow\infty$ oscillations vanish as the
  band-limited Delta converges to $\dirac_s(\rad) = \rad^{-2}
  \dirac^\reals(\rad-s)$. }
\label{fig:deltas}
\end{figure}

In order to recover a real space representation of the radial
translation operator we must first consider the Dirac delta function
on the radial line.  We define the Dirac delta on the radial line at
position $s$ by $\dirac_s(\rad) \equiv \rad^{-2} \dirac^\reals(\rad-s)$,
where $\dirac^\reals$ is the usual Dirac delta defined on the infinite
line $\reals$. The Dirac delta on the radial line satisfies the
following normalisation and sifting properties, respectively:
\begin{align}
\int_{\reals^+} \dx\rad \rad^2 \dirac_s(\rad) = 1 ; \\
\int_{\reals^+} \dx\rad \rad^2 f(\rad) \dirac_s(\rad) = f(s) .
\end{align}
The harmonic expansion of the Dirac delta is given by
\begin{equation}
  \label{eqn:dirac_bl}
  \dirac_{s}(\rad) = 
  \sum_{\ip = 0}^{\infty}
  K_\ip(s) K_{\ip} (\rad) ,
\end{equation}
which follows trivially by the sifting property. For the analysis of
band-limited functions, it is sufficient to consider the band-limited
Dirac delta (see \fig{\ref{fig:deltas}}), where the summation of
\eqn{\ref{eqn:dirac_bl}} is truncated to $P-1$.

With the Dirac delta function now defined on the radial line, we show
that the radial translation operator defined above is simply the
convolution of a function with the shifted Dirac delta function:
\begin{equation}
  \label{eqn:translation_radial_real}
  (f \star \dirac_s)(\rad)
  =  \sum_{\ip = 0}^{\infty}
  f_{\ip}  K_\ip(s) K_{\ip} (\rad) 
  = (\transl_{s}f)(\rad) ,
\end{equation}
where the final equality follows by
\eqn{\ref{eqn:translation_radial_harmonic_expansion}}.  Radial
convolution and translation are thus the natural analogues of the
respective operators defined on the infinite line. 

We define the translation operator on the ball by combining the
angular and radial translation operators, giving
\begin{equation}
  \transl_{\vect{\rad}} \equiv \transl_{\rad} \rot_{(\sas)} .
\end{equation}
The action of the radial translation operator on functions defined on
the ball is shown in \fig{\ref{fig:wavtransl}}.  The convolution on
the ball of $f \in \ltwo(\ball)$ with an axisymmetric kernel $h \in
\ltwo(\ball)$ is then defined by the inner product
\begin{equation}
  \label{eqn:convolution_ball_real}
  (f \star h)(\vect{r}) 
  \equiv \langle f | \transl_{\vect{\rad}} h\rangle_\ipball
  = \int_{\ball} \dx^3\vect{s} f(\vect{s}) (\transl_{\vect{\rad}} h)^\cconj (\vect{s}),
\end{equation}
where $\vect{s} \in \ball$. 
In harmonic space, axisymmetric convolution on the ball may be written 
\begin{equation}
  \label{eqn:convolution_ball_harmonic}
  {(f \star h)}_{\el \m \ip}
  =  \langle f \star h | Z_{\el \m \ip} \rangle_\ipball 
  =  \sqrt{ \frac{4\pi}{2\el+1}} {f}_{\el \m \ip} {h}^*_{\el 0 \ip} ,
\end{equation}
with $f_{\el \m \ip} = \langle f|Z_{\el \m \ip} \rangle_\ipball$ and $h_{\el 0
  \ip}\delta_{\m 0} = \langle h|Z_{\el \m \ip} \rangle_\ipball$.

% -------------------------------------------------------------------------
\section{Flaglets on the ball}
\label{sec:flaglet}

With an exact harmonic transform and a convolution operator defined on
the ball in hand, we are now in a position to construct our exact
wavelet transform on the ball, which we call the flaglet transform
(for Fourier-LAGuerre wavelet transform) \cite{leistedt:flaglets}. 

For a function of interest $ f\in \ltwo(\ball)$, we define its
\mbox{$jj^\prime$-th} wavelet coefficient $W^{\Psi^{jj^\prime}}\in
\ltwo(\ball)$ by the convolution of $f$ with the axisymmetric wavelet,
or flaglet, $\Psi^{jj^\prime}\in \ltwo(\ball)$:
\begin{equation}
  W^{\Psi^{jj^\prime}}(\vect{\rad})  \equiv (f \star
  \Psi^{jj^\prime})(\vect{\rad}) 
  = \langle f | \transl_{\vect{\rad}}  \Psi^{jj^\prime} \rangle_\ipball.
\end{equation} 
The scales $j,j^\prime\in \naturals_0^+$ respectively relate to angular and
radial spaces. 
The wavelet coefficients contain the detail information of the signal
only; a scaling function and corresponding scaling coefficients must
be introduced to represent the low-frequency, approximate information
of the signal.  The scaling coefficients $W^\Phi \in \ltwo(\ball)$ are
defined by the convolution of $f$ with the scaling function $\Phi\in
\ltwo(\ball)$:
\begin{equation}
  W^\Phi(\vect{\rad}) \equiv (f \star \Phi)(\vect{\rad}) 
  = \langle f | \transl_{\vect{\rad}} \Phi \rangle_\ipball.
\end{equation}

Provided the flaglets and scaling function satisfy an admissibility
property (defined below), the function $f$ may be reconstructed exactly from its wavelet
and scaling coefficients by
\begin{equation}
\begin{split}
  f(\vect{\rad}) 
  = &\int_{\ball} \dx^3\vect{\rad}^\prime 
  W^{\Phi}(\vect{\rad}^\prime)(\transl_{\vect{\rad}} \Phi)(\vect{\rad}^\prime) \\
  &+ \sum_{j=J_0}^{J} \sum_{j^\prime=J^\prime_0}^{J^\prime}  
  \int_{\ball} \dx^3\vect{\rad}^\prime 
  W^{\Psi^{jj^\prime}}(\vect{\rad}^\prime)(\transl_{\vect{\rad}} \Psi^{jj^\prime})(\vect{r}^\prime).
\end{split}
\end{equation}
The parameters $J_0$ and $J$ ($J^\prime_0$ and $J^\prime$) define the
minimum and maximum wavelet scales considered respectively for the
angular (radial) space and depend on the band-limit of $f$ and the
specific definition of the wavelets and scaling function (see
\cite{leistedt:flaglets}).

\begin{figure}
\centering
\subfigure[$ \Psi^{jj^\prime}(\vect{\rad})$ translated by $r=0.2$]{\includegraphics[trim= 1cm 24.4cm 0.5cm 1cm, clip, width=3.5cm]{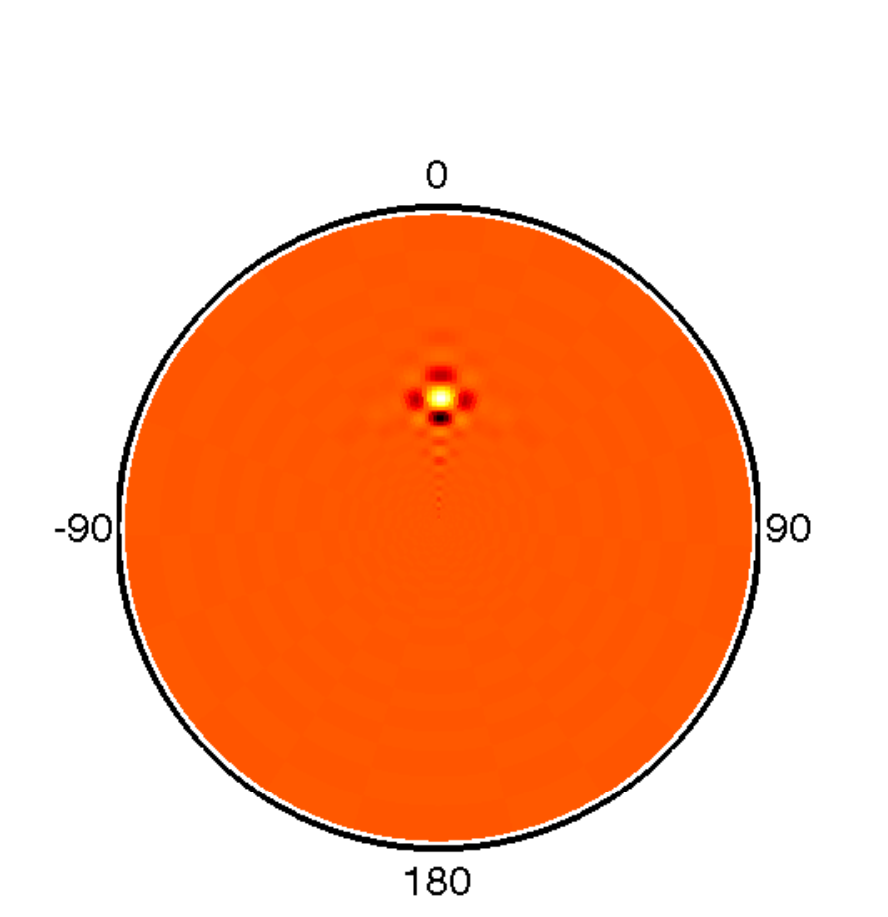}\includegraphics[trim= 6cm 23.4cm 1cm 1cm, clip, width=3.5cm]{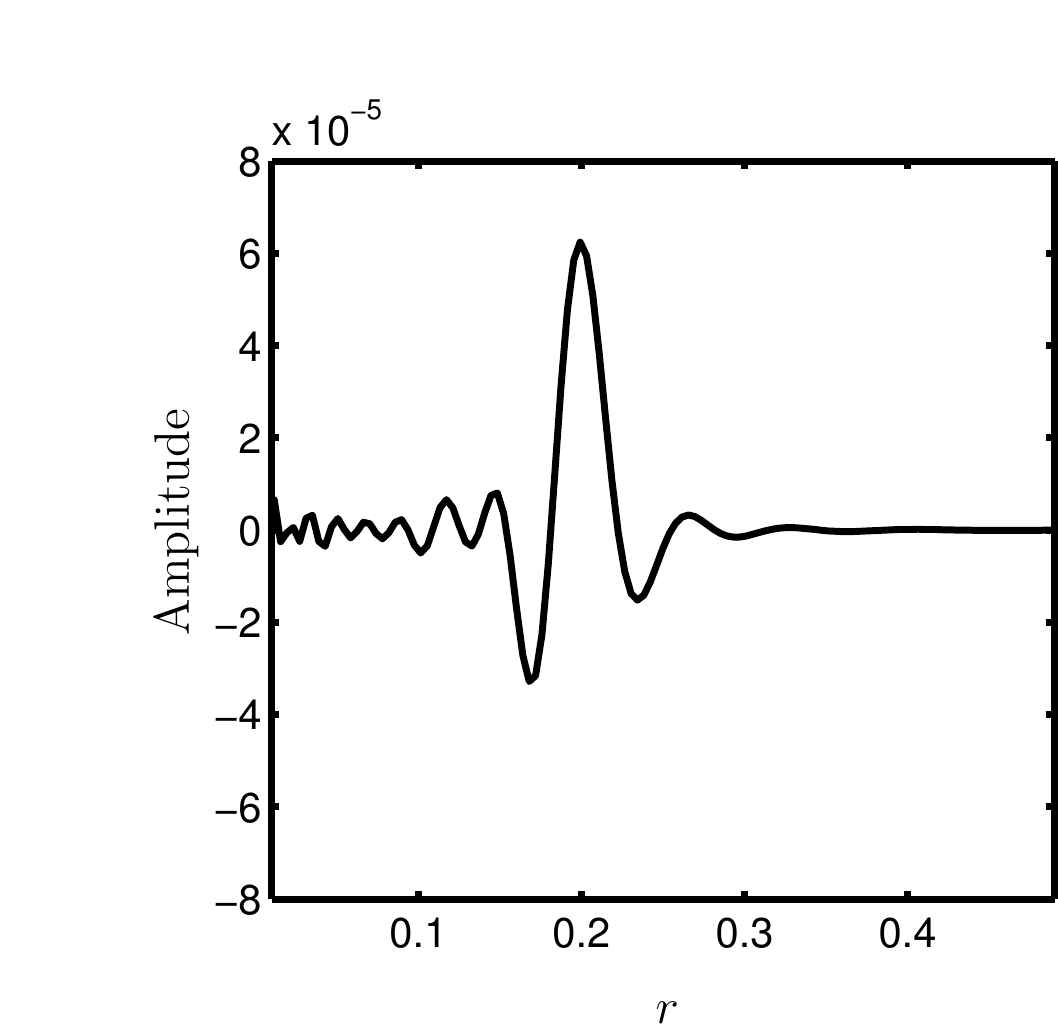}}\\
%\subfigure[Wavelet kernel translated by $r=0.3$]{\includegraphics[trim= 1cm 14cm 0.5cm 12.5cm, clip, width=3.5cm]{translationoperator_bis.pdf}\includegraphics[trim= 6cm 12.9cm 1cm 12.4cm, clip, width=3.5cm]{translationoperator.pdf}}\\
\subfigure[$ \Psi^{jj^\prime}(\vect{\rad})$ translated by $r=0.4$]{\includegraphics[trim= 1cm 3.6cm 0.5cm 22.8cm, clip, width=3.5cm]{figures/translationoperator_bis}\includegraphics[trim= 6cm 2.6cm 1cm 22.8cm, clip, width=3.5cm]{figures/translationoperator}}
\caption{Slices of the flaglet $ \Psi^{jj^\prime}(\vect{\rad})$ with
  $j=j^\prime=5$ constructed on the ball of radius $R=1$ at resolution
  $P=L=64$. The three-dimensional flaglets can be visualised by
  rotating the slices of the left panel (zoomed on a ball of radius
  $r=0.5$ for clarity) around the vertical axis passing through the
  origin. The radial profiles are shown in the right panels. Flaglets
  are well localised in both real and Fourier-Laguerre spaces.
  Furthermore, their angular aperture is invariant under radial
  translation.}
\label{fig:wavtransl}
\end{figure}

The admissibility condition under which a band-limited
function $f$ can be reconstructed exactly is given by
the following resolution of the identity:
\begin{equation}
  \label{eqn:admissibility}
  \frac{4\pi}{2\ell+1} 
  \Biggl( |{\Phi}_{\el 0 \ip}|^2 + \sum_{j=J_0}^{J}
  \sum_{j^\prime=J^\prime_0}^{J^\prime}  
  |{\Psi}^{jj^\prime}_{\el 0 \ip}|^2 \Biggr) = 1, \quad \forall \ell, p ,
\end{equation}
where ${\Phi}_{\el 0 \ip}\delta_{\m 0} = \langle \Phi|Z_{\el \m \ip}
\rangle_\ipball$ and ${\Psi}^{jj^\prime}_{\el 0 \ip}\delta_{\m 0} = \langle
\Psi^{jj^\prime}|Z_{\el \m \ip} \rangle_\ipball$. We refer the reader to our
previous article \cite{leistedt:flaglets} for an example of the
construction of specific wavelets and scaling functions that satisfy
the admissibility condition, where we construct suitable wavelets by
tiling the $\el$-$\ip$ harmonic plane.  The resulting wavelets are
plotted in \fig{\ref{fig:wavtransl}}.

We prove that flaglets are a tight frame by showing they satisfy
\begin{eqnarray}
\label{eqn:frame_property}
%\begin{split}
  &&A \| f \|^2_{\ipball} 
  \leq
  \int_{\ball} \dx^3\vect{\rad}
  |\langle f | \transl_{\vect{\rad}}{\Phi} \rangle_\ipball |^2\\
  && \:\: +
  \sum_{j=J_0}^{J}
  \sum_{j^\prime=J^\prime_0}^{J^\prime} 
  \int_{\ball} \dx^3\vect{\rad}
  |\langle f | \transl_{\vect{\rad}}{\Psi}^{jj^\prime} \rangle_\ipball |^2
  \leq
  B \| f \|^2_{\ipball} , \nonumber
%\end{split}
\end{eqnarray}
with $A=B\in\realsnz$, for any band-limited $f \in \ltwo(\ball)$, and
where $\| \cdot \|_{\ipball}^2 \equiv \langle \cdot | \cdot
\rangle_\ipball$.  We adopt a shorthand integral notation in
\eqn{\ref{eqn:frame_property}}, although by appealing to our
exact quadrature rule these integrals may be replaced by finite sums.
Noting the harmonic expression for axisymmetric convolution given by
\eqn{\ref{eqn:convolution_ball_harmonic}} and the orthogonality of the
Fourier-Laguerre basis functions, it is straightforward to show that
the term of \eqn{\ref{eqn:frame_property}} bounded between
inequalities may be written
\begin{equation}
\begin{split}
  \sum_{\ip = 0}^{\ipmax-1}\sum_{\el = 0}^{\elmax-1}\sum_{\m = -\el}^{\el} 
  \frac{4\pi}{2\el+1} 
  \Biggl (&
  |{\Phi}_{\el 0 \ip}|^2 
  |f_{\el \m \ip}|^2 \\
  &+
  \sum_{j=J_0}^{J}
  \sum_{j^\prime=J^\prime_0}^{J^\prime} 
  |{\Psi}^{jj^\prime}_{\el 0 \ip}|^2
  |f_{\el \m \ip}|^2 
  \Biggr ) \\
  = 
  \sum_{\ip = 0}^{\ipmax-1}\sum_{\el = 0}^{\elmax-1}\sum_{\m =-\el}^{\el} 
  |f_{\el \m \ip}|^2 
  &= 
  \int_{\ball} \dx^3\vect{\rad}
  |f(\vect{\rad})|^2 
  = \| f \|^2_{\ipball},
\end{split}
\end{equation}
where the second line follows from the admissibility property
\eqn{\ref{eqn:admissibility}}.  Thus, we find flaglets indeed
constitute a tight frame with $A=B=1$, implying the energy of $f$ is
conserved in flaglet space.

We have developed the public {\tt
  FLAGLET}\footnote{\label{flagletwebsite}\url{http://www.flaglets.org/}}
code \cite{leistedt:flaglets} to compute the flaglet transform. The
{\tt FLAGLET} code computes the exact forward and inverse flaglet
transform at machine precision, exploiting a fast multiresolution
algorithm, and is stable to extremely large band-limits (the
computation time and numerical precision of the {\tt FLAGLET} code is
evaluated in detail in \cite{leistedt:flaglets}, where a toy
application is also presented).  {\tt FLAGLET} relies
on the public code
{\tt S2LET}\footnote{\url{http://www.s2let.org/}}
\cite{leistedt:s2let_axisym}
(to compute wavelet transforms on the sphere),
{\tt FLAG}\footnoteref{flagletwebsite}
\cite{leistedt:flaglets},
{\tt SSHT}\footnote{\url{http://www.jasonmcewen.org/}}
\cite{mcewen:fssht}
and 
{\tt FFTW}\footnote{\url{http://www.fftw.org/}},
and supports both the C and Matlab programming
languages.  

To summarise, flaglets live naturally on the ball (with an angular
opening that is invariant under radial translation), yield a
theoretically exact wavelet transform on the ball (in both the
continuous and discrete settings), and exhibit a fast multiresolution
algorithm.  It is our hope that flaglets will prove useful for
analysing data defined on the ball.  Indeed, in the near future we
intend to apply flaglets to study the large-scale structure of the
Universe traced by the distribution of galaxies.

\section*{Acknowledgements}

JDM is supported by a Newton International Fellowship from the Royal
Society and the British Academy. BL is supported by the Perren Fund
and the IMPACT Fund.

% -------------------------------------------------------------------------
\bibliographystyle{plain}
%\footnotesize
\bibliography{bib}

\end{document}

%% file: macros.tex
\newcommand{\eqn}[1]{(#1)}

\newcommand{\fig}[1]{Fig.~#1}

\newcommand{\eg}{\mbox{\it e.g.}}
\newcommand{\ie}{\mbox{\it i.e.}}

\newcommand{\cconj}{\ensuremath{\ast}} 
 
\newcommand{\reals}{\ensuremath{\mathbb{R}}}
\newcommand{\realsnn}{\ensuremath{\mathbb{R^+}}}
\newcommand{\realsnz}{\ensuremath{\mathbb{R}^{+}_{\ast}}}
\newcommand{\integers}{\ensuremath{\mathbb{Z}}}
\newcommand{\naturals}{\ensuremath{\mathbb{N}}}

\newcommand{\ltwo}{\ensuremath{\mathrm{L}^2}}
\newcommand{\sphere}{\ensuremath{{\mathbb{S}^2}}}
\newcommand{\sothree}{\ensuremath{{\mathrm{SO}(3)}}}

\newcommand{\ball}{{\ensuremath{\mathbb{B}^3}}}
\newcommand{\ipball}{\ensuremath{\ball}}
\newcommand{\ipsphere}{\ensuremath{\sphere}}
\newcommand{\iphalfline}{\ensuremath{\realsnn}}
\newcommand{\vect}[1]{\ensuremath{\mbox{\boldmath ${#1}$}}}

\newcommand{\dx}{\ensuremath{\mathrm{\,d}}}
\newcommand{\dmu}[1]{\ensuremath{\dx \Omega(#1)}}

\newcommand{\saa}{\ensuremath{\theta}}
\newcommand{\sab}{\ensuremath{\varphi}}
\newcommand{\sas}{\ensuremath{\saa, \sab}}

\newcommand{\euls}{\ensuremath{\eula, \eulb, \eulc}}
\newcommand{\eula}{\ensuremath{\alpha}}
\newcommand{\eulb}{\ensuremath{\beta}}
\newcommand{\eulc}{\ensuremath{\gamma}}
\newcommand{\el}{\ensuremath{\ell}}
\newcommand{\m}{\ensuremath{m}}

\newcommand{\ip}{\ensuremath{p}}

\newcommand{\elmax}{\ensuremath{{L}}}
\newcommand{\ipmax}{\ensuremath{{P}}}
\newcommand{\p}{\ensuremath{^\prime}}

\newcommand{\rad}{\ensuremath{r}}

\newcommand{\kron}[2]{\ensuremath{\delta_{{#1}{#2}}}}

\newcommand{\dirac}{\ensuremath{\delta}}

\renewcommand{\exp}[1]{\ensuremath{{\rm e}^{#1}}}
\newcommand{\shfarg}[3]{\ensuremath{Y_{#1#2}({#3})}}

\newcommand{\shf}[2]{\ensuremath{Y_{#1#2}}}

\newcommand{\shc}[3]{\ensuremath{{#1}_{{#2}{#3}}}}
\newcommand{\shcc}[3]{\ensuremath{{#1}_{{#2}{#3}}^\cconj}}

\newcommand{\rot}{\ensuremath{\mathcal{R}}}

\newcommand{\transl}{\ensuremath{\mathcal{T}}}